\documentclass[12pt,preprint]{aastex}
\shorttitle{Two new accreting pulsating white dwarf stars}
\shortauthors{Mukadam et al.}
\newcommand{\etal}{et al.}

\newcommand{\zzc}{ZZ~Ceti}
\newcommand{\cvzA}{SDSS\,J074531.91+453829.5}
\newcommand{\cvzB}{SDSS\,J091945.10+085710.0}
\newcommand{\cvza}{SDSS\,0745+4538}
\newcommand{\cvzb}{SDSS\,0919+0857}
\newcommand{\novA}{SDSS\,J171145.08+301320.0}
\newcommand{\nova}{SDSS\,1711+3013}
\newcommand{\msun}{M_{\odot}}

\begin{document}

\title{Discovery of Two New Accreting Pulsating White Dwarf Stars}
\author{Anjum S. Mukadam\altaffilmark{1,2,3}, B. T. G\"{a}nsicke\altaffilmark{4}, P. Szkody\altaffilmark{1,2},
A. Aungwerojwit\altaffilmark{4,5}, Steve B. Howell\altaffilmark{6,7}, O. J. Fraser\altaffilmark{1,2}, and N. M. Silvestri\altaffilmark{1,2}}
\altaffiltext{1}{Department of Astronomy, University of Washington, Seattle, WA\,-\,98195-1580, USA}
\altaffiltext{2}{Hubble fellow; anjum@astro.washington.edu}
\altaffiltext{3}{Visiting Astronomer, Apache Point Observatory, which is owned and operated by the Astrophysical Research Consortium} 
\altaffiltext{4}{Department of Physics, University of Warwick, Coventry, CV4 7AL, UK}
\altaffiltext{5}{Department of Physics, Faculty of Science, Naresuan University, Phitsanulok, 65000, Thailand}
\altaffiltext{6}{WIYN Observatory and National Optical Astronomy Observatory, 950 North Cherry Avenue, Tucson, AZ\,-\,85719, USA}
\altaffiltext{7}{Visiting Astronomer, Kitt Peak National Observatory, National Optical Astronomy Observatories, which is operated by the Association of Universities for Research in Astronomy, Inc. (AURA) under cooperative agreement with the National Science Foundation. The WIYN Observatory is a joint facility of the University of Wisconsin-Madison, Indiana University, Yale University, and the National Optical Astronomy Observatories.}

\begin{abstract}
We report the discovery of two new accreting pulsating white dwarf stars amongst the cataclysmic variables
of the Sloan Digital Sky Survey: \cvzA\ and \cvzB. We observe high amplitude non-sinusoidal variations
of 4.5--7\% at a period close to 1230\,s in the optical light curves of \cvzA\ and
a low amplitude variation of 0.7--1.6\% near 260\,s in the light curves of \cvzB.
We infer that these optical variations are a consequence of nonradial g-mode pulsations
in the accreting primary white dwarfs of these cataclysmic variables.
However we cannot rule out the remote possibility that the 260\,s period
could be the spin period of the accreting white dwarf \cvzB.
We also uncovered a non-variable \novA\ during our search;
our two observing runs exclude any pulsation related periodicities in the range of
85--1400\,s with an amplitude $\geq$0.5\%.
This discovery paper brings the
total number of known accreting white dwarf pulsators to eleven.
\end{abstract}

                                                                                                         
\section{Introduction}
Cataclysmic variables (CVs) are close binary systems in which a late-type
star (secondary) fills its Roche lobe and transfers mass through
the inner Lagrangian point to a white dwarf (primary).
GW Librae was the first CV discovered with an accreting
white dwarf
showing photometric variations
consistent with nonradial g-mode pulsations observed in non-interacting white 
dwarf stars \citep{WarneravanZyl98,vanZylet00,vanZylet04}.
This discovery has opened a new venue of opportunity for us
to learn about the stellar parameters of accreting variable white dwarfs
using asteroseismic techniques. We can probe the insides of a white dwarf using asteroseismology
just like seismologists on Earth can learn about the interior of our planet using earthquakes.
A unique model fit to the observed periods of the variable
white dwarf can reveal information about the stellar mass, core
composition, age, rotation rate, magnetic field strength, and
distance \citep[see the review paper][]{Winget98}.
This information could prove vital in determining the system
dimensions and parameters, and may be the only way to derive
conclusive results for non-eclipsing CVs. Pulsating white dwarfs in CVs are
also interesting to the variable white dwarf community because they can
help establish the effect of accretion on pulsations.

We report new accreting pulsators, \cvzA\ and \cvzB\ (hereafter \cvza\ \& \cvzb),
originally discovered to be CVs in the Sloan
Digital Sky Survey (SDSS) by \citet{Szkodyet06} and \citet{Szkodyet05} respectively.
This result comes from our long-term
search to discover additional accreting pulsating white dwarf stars in CVs.
There are now eleven accreting pulsating white dwarfs known,
including the two presented in this paper \citep[see][]{vanZylet04,WoudtaWarner04,WarneraWoudt04,
Pattersonet05a,Pattersonet05b,Vanlandinghamet05,Araujo-Betancoret05,Gaensickeet06,Nilssonet06}.
Our goal is to
establish the pulsationally unstable region(s) for accreting white dwarfs, which requires
a statistically significant number of these systems with well-determined temperatures. Finding the
non-variables around an instability strip is equally important in determining its boundaries as discovering
variables close to its edges. With this context in mind, we also report a
non-variable \novA\ (hereafter \nova), discovered as a CV by \citet{Szkodyet04}. We 
observed the system twice and place useful non-variability limits related to the absence of
pulsation in the primary white dwarf.                                                                                                            

\section{Selection of Candidates}
We select those CVs for high-speed photometry whose optical spectra include prominent broad absorption lines
from the white dwarf as well as the contaminating narrow emission features from the accretion disk.
The absorption lines indicate that the white dwarf flux dominates the light from the CV.
While this is not a {\it sufficient} criterion for pulsation in the primary white dwarf, 
these are the only systems where the search is even possible.
When the orbital period of a CV is $\sim$80-90\,min, it is
near the evolutionary period minimum and the CV exhibits the
smallest rate of mass transfer. \citet{Gaensickeet06} used the optical accretion luminosity 
to constrain the rate of mass transfer to be greater than $\sim10^{-13}\msun/yr$, while \citet{KolbaBaraffe99}
and \citet{Howellet97} theoretically compute the rate of mass transfer at the period minimum to
be few times $\sim10^{-11}\msun/yr$. We can
expect that the low rates of mass transfer allow a larger surface area of the
white dwarf to be visible, making CVs near the period minimum ideal candidates
to find additional accreting pulsators.
Note that the two criteria mentioned here are not independent. 
                                                                                                                   
\citet{Arraset06} investigate the temperature range in which
models of accreting white dwarfs with a wide range of masses and Helium enrichment from the donor star
would be pulsationally unstable. They find a H/HeI instability strip for
accreting model white dwarfs with a blue edge near $\leq$12000\,K for a 0.6\,$\msun$ star.
The blue edge shifts to hotter (cooler) temperatures by about 2000\,K for a factor of 10
increase (decrease) in gravity; we can expect the blue edge at 14000\,K corresponding to $\log~g$\,=\,9
and at 10000\,K for $\log~g$\,=\,7.
This theoretical instability strip is similar to the \zzc\ instability strip\footnote{Non-interacting hydrogen atmosphere (DA) white dwarfs are observed to
pulsate in a narrow instability strip located within the temperature range
10800--12300\,K for $\log~g\approx 8$ \citep{Bergeronet95,Bergeronet04,
KoesteraAllard00,KoesteraHolberg01,Mukadamet04b,Gianninaset05}, and
are also known as the \zzc\ stars.}.
For accreting model white dwarfs with a
high He abundance ($>$\,0.38), \citet{Arraset06} find an additional hotter instability strip
at $\approx$15000\,K due to HeII ionization. The boundaries of this
intermediate instability strip depend on the Helium abundance and
the mass of the accreting white dwarf. For a He abundance higher than 0.48, these theoretical
instability strips essentially merge. \citet{Arraset06} expect that there are thus
two distinct instability strips for accreting model white dwarfs with a He abundance
between about 0.38 and 0.48.

Optical spectra do not lead to sufficiently reliable temperatures for the white dwarf due to
contamination from the accretion disk. Accurate temperature determination requires ultraviolet
spectra \citep[see][]{Szkodyet07} that we do not have.
Hence we are unable to impose any
additional selection criteria that depend on the temperature of the primary white dwarf,
despite the theoretical framework mentioned above. Besides,
an empirical instability strip for accreting white dwarfs has not yet been established. There
are only five accreting pulsators with well determined temperatures from
ultra-violet spectra; GW Librae, SDSS\,013132.39-090122.3, SDSS\,J161033.64-010223.3,
and SDSS\,J220553.98+115553.7 lie within the range 
14500--15000\,K \citep{Szkodyet02,Szkodyet07}, while \citet{Araujo-Betancoret05}
find the accreting white dwarf pulsator HS\,2331+3905 to be at 10500\,K.

\section{Modeling the SDSS spectra}
Although we do not use temperatures from optical spectra to select candidates, we
estimated temperatures for the two pulsators and the non-pulsator discovered during our search.
In order to establish the white dwarf temperature, the distance to the
system, and to constrain the spectral type of the donor star, we
model the optical SDSS spectra as the sum of a white dwarf, an
accretion disk, and a late-type secondary star. For the white dwarf, we
assume a surface gravity of $\log g=8$, corresponding to a mass of
$M_\mathrm{wd}\simeq0.6M_\odot$ and a radius of
$R_\mathrm{wd}=8.7\times10^8$\,cm.
The contaminating emission from the accretion disk
fills in the Balmer absorption from the white dwarf, making it impossible to determine
$\log$~g independently.
We then compute a grid of pure
hydrogen model spectra covering effective temperatures in the range
8000--20000\,K using the codes TLUSTY/SYNSPEC
\citep{HubenyaLanz95,LanzaHubeny95}. We model the accretion disk as
an isothermal/isobaric slab of hydrogen \citep{Gaensickeet99}.
For the donor star, we use the M-dwarf
templates of \citet{Beuermannet98} and L-dwarf
templates from \citet{Kirkpatricket99,Kirkpatricket00}.
We fix the radius of the secondary star to
$R_2=8.6\times10^9$\,cm, which corresponds to a stellar mass of 
$M_2=0.08M_\odot$, close to the dividing line between
main-sequence and sub-stellar objects, at an orbital period of
80\,min.

\begin{figure}[p]
\centering
\includegraphics[height=6.5in,clip=true]{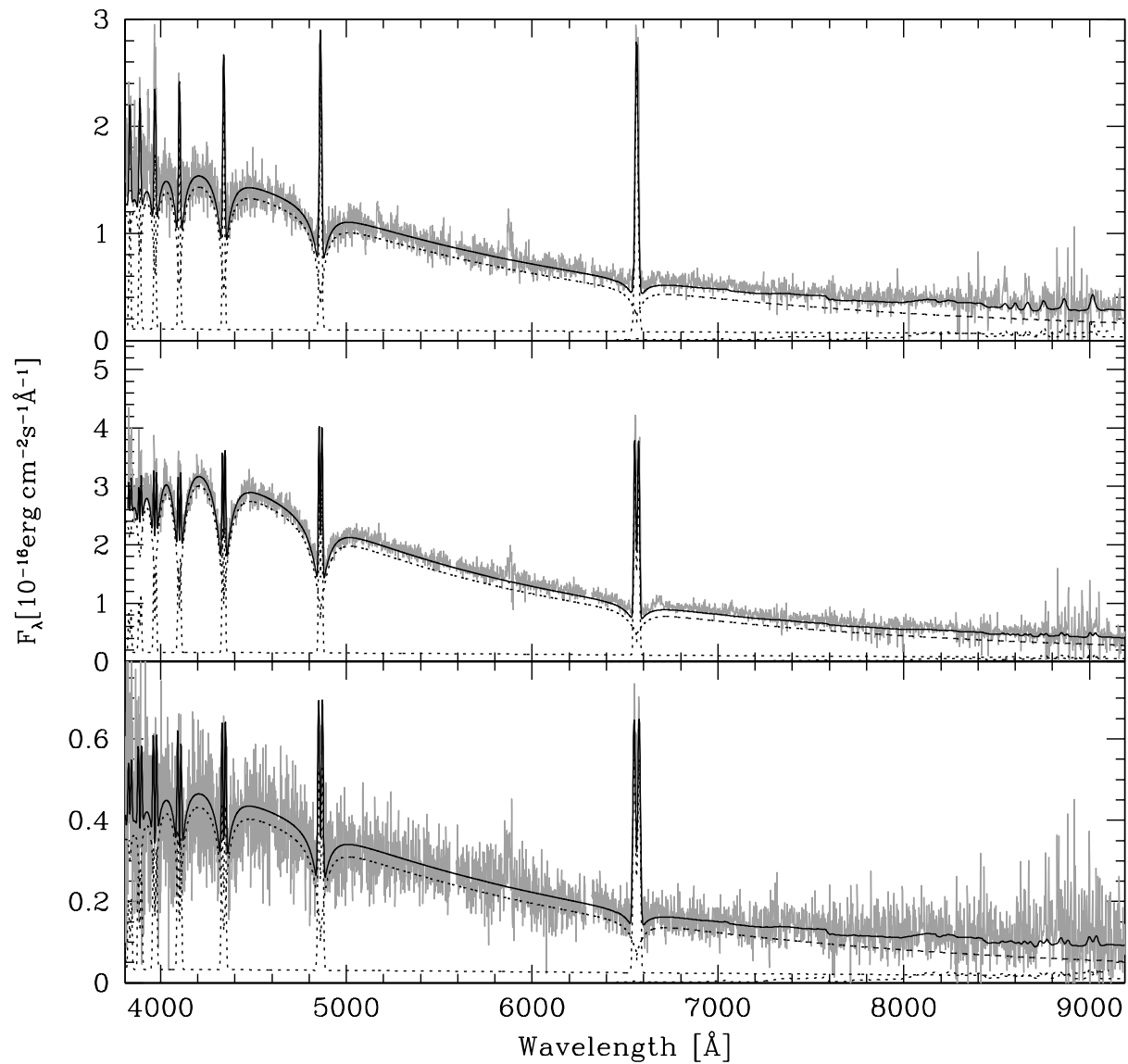}
\caption{\label{t-fit} We show the observed optical spectra of \cvza\ (top panel), \cvzb\ (middle panel),
and \nova\ (bottom panel) in gray along with their respective model fits (black lines)
that include three components (dotted lines): primary white dwarf, accretion disk with emission lines,
and the secondary star.}
\end{figure}

The free parameters of our three-component model are the white
dwarf temperature $T_\mathrm{wd}$, the distance $d$, the spectral
type of the secondary star Sp(2), the disc temperature
$T_\mathrm{d}$, and its surface density $\Sigma_\mathrm{d}$.
Considering the number of free parameters, we refrain from applying
a formal $\chi^2$ fit and model the SDSS spectrum as follows.
We initially calculate a disk spectrum for the
given choice of $T_\mathrm{d}$ and $\Sigma_\mathrm{d}$, and scale the
H$\alpha$ emission line fluxes of the model to the observed
value. Next, for the chosen $T_\mathrm{wd}$, we adjust $d$ to
reproduce the observed flux, and judge the goodness of the fit by eye. We ensure that the
combination of the white dwarf and the disk fits the Balmer absorption lines
in the SDSS spectrum, as well as the slope of the continuum at
$\lambda<5000$\,\AA, where the contribution of the donor star can
safely be ignored. Finally, we scale the selected M/L dwarf template
for $d$ as determined from the white dwarf fit. The absence of
noticeable TiO bands in the SDSS spectrum provides an {\it early} 
limit on the spectral type of the donor~--~{\it later} spectral types cannot
be excluded. We list the parameters corresponding to our model fits to the SDSS
spectra in Table\,\ref{t-fitparameter}, and show these models along with the observed
spectra in Figure\,\ref{t-fit}.
The white dwarf temperatures estimated
from the optical spectra are uncertain by $\pm$1000--2000\,K.

\clearpage
\begin{deluxetable}{llllll}
\tabletypesize{\footnotesize}
\tablecolumns{5}
\tablewidth{0pc}
\tablecaption{\label{t-fitparameter}
Estimated system parameters for \cvza, \cvzb, and \nova\
determined from a three-component (white dwarf, accretion disk, donor star)
model fit to their optical SDSS spectra.}
\tablehead{
\colhead{CV} &
\colhead{$T_\mathrm{wd}$\,[K]} &
\colhead{Sp(2)} &
\colhead{$d$\,[pc]} &
\colhead{$T_\mathrm{d}$\,[K]} &
\colhead{$\Sigma_\mathrm{d}\,[\mathrm{g\,cm^{-2}}]$}}
\startdata
\cvza\ & 11000 & $>$M8 & 280 & 6600 & $1.7\times10^{-2}$ \\
\cvzb\ & 13000 & $>$M9 & 235 & 6300 & $1.5\times10^{-2}$ \\
\nova\ & 10500 & $>$M8 & 480 & 6500 & $1.4\times10^{-2}$\\
\enddata
\end{deluxetable}
\clearpage
                                                                                                         
\clearpage                                                                                                         
                                                                                                         
The parameters of all three systems discussed here are similar to
those of HS\,2331+3905 \citep{Araujo-Betancoret05}, SDSS\,1339+4847
\citep{Gaensickeet06}, and SDSS\,1035+0551 \citep{Southworthet06},
which were analyzed in a similar fashion,
i.e. white dwarf temperatures in the range 10500--13000\,K, donor
stars later than M8, disk temperatures around 6500\,K, and surface
densities of a few $10^{-2}\mathrm{g\,cm^{-2}}$. Follow-up observations
of SDSS\,1035+0551 revealed eclipses in the light curve, confirming the existence of
a brown dwarf donor with $0.052\pm0.002M_\odot$
\citep{Littlefairet06}. The limits on the
spectral types determined from the SDSS spectra suggest that
a substantial fraction of these objects could harbor brown dwarf secondary stars.

\section{Observations and Data Reduction}
We obtained optical high-speed time-series photometry on suitable SDSS CVs
using the 3.5\,m telescope at 
Apache Point Observatory (APO) in New Mexico, the 
2.2\,m telescope at Calar Alto Observatory (CAO) in Spain, and the
3.5\,m WIYN telescope at Kitt Peak National Observatory in Arizona.

We used the Dual Imaging Spectrograph (DIS) in imaging mode, mounted at the Nasmyth focus 
of the 3.5\,m telescope at APO. The instrument design
utilizes a dichroic that splits the beam of white light at 555\,nm. The blue and red portions
of the beam are then incident on two different CCD cameras. We present filterless light curves obtained
using the blue CCD camera in this paper, sensitive in the wavelength range 350-555\,nm.
The instrument has a read noise of about 4.6 electrons RMS, a gain of 1.75 e/ADU, and a plate
scale of 0.4 arcsec/pixel at the 3.5\,m telescope at APO.
We used windowing to read a small portion of the CCD in order
to reduce the read time to 2--4\,s. Including an additional overhead of 5\,s from 
CCD flushing, writing data to disk, and other software delays, we obtained a dead time of order 7--9\,s between 
exposures. We used a standard IRAF reduction to extract sky-subtracted light curves 
from the CCD frames using weighted circular aperture photometry \citep{O'Donoghueet00}.

At the 2.2\,m CAO telescope, we used the Calar Alto Faint Object Spectrograph (CAFOS)
to obtain filterless CCD photometry.
We used windowing and binning in order
to reduce the read time and hence the dead time between exposures. We applied a bias
correction and flat-fielded the images within MIDAS, and then used the
Source-extractor algorithm \citep{BertinaArnouts96} to perform
aperture photometry on all objects in the field-of-view and determined corresponding light curves.

We obtained the photometric observations at the WIYN 3.5\,m telescope
using the Orthogonal Parallel Transfer Imaging camera (OPTIC),
the prototype orthogonal transfer CCD imager \citep{Tonryet02}. It
consists of two 2K\,x\,4K orthogonal transfer CCDs mounted together in a single
dewar. The camera has a
read noise of $<$4 electrons RMS, a gain of 1.45 e/ADU, and a plate
scale of 0.12 arcsec/pixel at the WIYN 3.5\,m telescope.
We used OPTIC in conventional mode \citep[see][]{Howellet03} and all the
time-series images were binned 2x2, providing an 8\,s readout time
between successive exposures.
We observed through a
blue bandpass BG-39 filter ($\sim$B+V) for our
observations. We reduced all the data from the WIYN telescope
in exactly the same manner as the APO data.

Although the existing instruments at all the telescopes mentioned above
allow short integration times, they are not entirely suitable
to study pulsating white dwarfs because they include a substantial dead time between exposures.
An ideal instrument such as a blue-sensitive frame transfer CCD camera would not only allow short exposure times,
but also include insignificant dead times between consecutive exposures, to enable data
acquisition with high time resolution.
We indicate our journal of observations acquired at the three different telescopes in Table\,\ref{jour-obs}.

\begin{deluxetable}{ccccc}
\tablecolumns{5}
\tablewidth{0pc}
\tablecaption{\label{jour-obs} Journal of Observations}\tablehead{
\colhead{Telescope}&\colhead{Object}&\colhead{Start Time (UTC)}& \colhead{Duration (hr)} & \colhead{Exposure (s)}}
\startdata
APO 3.5\,m & \cvza\ & 14 Oct 2005 11:29:10.458 & 0.83 & 15\\
APO 3.5\,m & \cvza\ & 30 Nov 2005 08:03:16.534 & 2.13 & 15\\
APO 3.5\,m & \cvzb\ & 01 Dec 2005 10:47:54.323 & 2.17 & 15\\
APO 3.5\,m & \cvzb\ & 05 Dec 2005 10:40:26.316 & 2.17 & 15\\
CAO 2.2\,m & \cvza\ & 20 Jan 2006 23:18:54.403 & 3.83 & 25\\
CAO 2.2\,m & \cvza\ & 21 Jan 2006 21:12:25.027 & 3.35 & 25\\
CAO 2.2\,m & \cvza\ & 22 Jan 2006 20:03:04.435 & 0.86 & 35\\
CAO 2.2\,m & \cvza\ & 23 Jan 2006 21:00:57.629 & 3.56 & 30\\
CAO 2.2\,m & \cvzb\ & 24 Jan 2006 03:26:17.578 & 2.95 & 30\\
APO 3.5\,m & \cvza\ & 30 Jan 2006 03:21:42.044 & 2.14 & 15\\
APO 3.5\,m & \cvzb\ & 20 Feb 2006 03:16:44.490 & 3.80 & 15\\
WIYN 3.5\,m & \nova\ & 29 May 2006 08:44:06.79 & 2.50 & 25\\
WIYN 3.5\,m & \nova\ & 30 May 2006 07:12:25.10 & 3.21 & 25\\
APO 3.5\,m & \cvzb\ & 16 Mar 2007 06:05:39.243 & 0.85 & 15\\
APO 3.5\,m & \cvzb\ & 18 Mar 2007 02:08:02.664 & 3.81 & 10\\
\enddata
\end{deluxetable}

After extracting the light curves, we divided the light curve of the target star with a sum of one or more 
comparison stars; we used brighter stars for the division whenever available as their light
curves have lower noise. After this preliminary reduction, we brought the
data to the same fractional amplitude
scale ($\Delta\,I/I$) and converted the mid-exposure times of the CCD images to Barycentric
Coordinated Time \citep[TCB;][]{Standish98}.
We then computed a Discrete Fourier Transform (DFT) for all the light curves
up to the Nyquist frequency.

\section{New accreting pulsating white dwarf \cvza}
\begin{figure}[!ht]
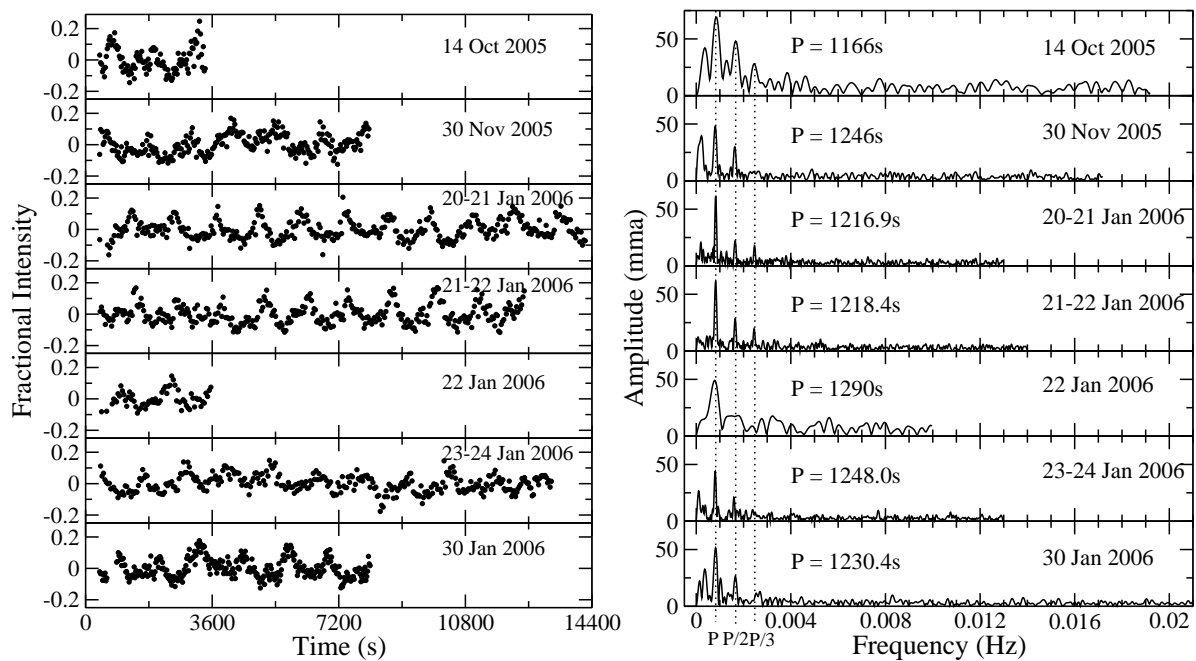

\centering
\includegraphics[width=3.2in,clip=true]{f2a}\includegraphics[width=3in,clip=true]{f2b}
\caption{\label{lcft-cvza} Light curves and Discrete Fourier Transforms (DFTs) of \cvza\ data obtained using the 3.5\,m telescope at 
Apache Point Observatory and the 2.2\,m telescope at Calar-Alto Observatory. We indicate the single
pulsation period $P$ observed in each run, along with the
first ($P$/2) and second harmonics ($P$/3), indicative of its nonlinear pulse shape.}
\end{figure}

We estimate an effective temperature of 11000\,K
for the accreting primary white dwarf in \cvza\ ($g$\,=\,19.05) using its optical SDSS spectrum
and assuming $\log$~g\,=\,8 (see Table\,\ref{t-fitparameter}).
We show the optical light curves and corresponding DFTs of \cvza\ data
obtained at the 3.5\,m telescope at APO and the 2.2\,m telescope at CAO in Figure\,\ref{lcft-cvza}.
We indicate the periodicities observed on individual nights in Table\,\ref{per-cvza}.

\subsection{Orbital Period}
The longest g-mode pulsation period observed in cool non-interacting
pulsating white dwarfs is $\sim$1400\,s. Hence we expect that
the photometric variations in the light curves with periods longer than
40\,min are indicative of the orbital period.
The 43.2\,min period observed on the 30th of January 2006 is merely a harmonic of the 86.5\,min
period observed on the 20th of January 2006. Using a weighted average of both these periodicities,
where the weights are inversely proportional to the 1\,$\sigma$ uncertainties,
we determine the longest observed period to be $86.3\,\pm\,2.0$\,min.
As the orbital periods of most of the
accreting pulsators fall in the range of 80-90\,min, indicative of their low rates of mass transfer (see section 2),
we conclude that our measurement of $86.3\,\pm\,2.0$\,min is the orbital period of \cvza.

\clearpage
\clearpage                                                                                                          

\begin{deluxetable}{lllll}
\tabletypesize{\footnotesize}
\tablecolumns{5}
\tablewidth{0pc}
\tablecaption{\label{per-cvza} We list all the periodicities observed in the light curves of \cvza.
We indicate the single pulsation mode in the accreting white dwarf in boldface and also list
the first and second harmonics of this dominant period, whenever present.} \tablehead{
\colhead{Observation} & \colhead{Orbital Period (min)} & \multicolumn{3}{c}{Pulsation Period (s); Amplitude (mma)\tablenotemark{\alpha}}\\
\colhead{Date (UTC)}& \colhead{Amplitude (mma)\tablenotemark{\alpha}}& \colhead{Fundamental P} & \colhead{First Harmonic P/2} & \colhead{Second Harmonic P/3}}
\startdata
14\,Oct\,2005 & &{\bf 1166\,$\pm$\,23; 70.4\,$\pm$\,6.9}& 597.4\,$\pm$\,9.2; 46.8\,$\pm$\,6.8 & \\
30\,Nov\,2005 & &{\bf 1246\,$\pm$\,11; 46.1\,$\pm$\,4.2}& 613.8\,$\pm$\,4.1; 28.2\,$\pm$\,4.3 & \\
20\,Jan\,2006 & 86.5\,$\pm$\,2.3; 20.0\,$\pm$\,2.7& {\bf 1216.9\,$\pm$\,2.7; 60.8\,$\pm$\,2.7} & 607.3\,$\pm$\,1.7; 24.0\,$\pm$\,2.7 & 405.4\,$\pm$\,1.0; 17.1\,$\pm$\,2.7\\
21\,Jan\,2006 & & {\bf 1218.4\,$\pm$\,3.4; 61.1\,$\pm$\,3.0} & 609.7\,$\pm$\,1.8;28.7\,$\pm$\,3.0 & 408.4\,$\pm$\,1.2; 18.9\,$\pm$\,3.0 \\
22\,Jan\,2006 & & {\bf 1290\,$\pm$\,40; 49.4\,$\pm$\,6.7} & & \\
23\,Jan\,2006 & & {\bf 1248.0\,$\pm$\,4.6; 44.6\,$\pm$\,3.2}&629.4\,$\pm$\,2.5; 21.0\,$\pm$\,3.2 & 418.1\,$\pm$\,2.1; 11.3\,$\pm$\,3.2\\
30\,Jan\,2006 & 43.2\,$\pm$\,0.8; 32.8\,$\pm$\,3.2\tablenotemark{\beta}& {\bf 1230.4\,$\pm$\,6.4; 54.2\,$\pm$\,3.2} & 611.0\,$\pm$\,3.1; 27.8\,$\pm$\,3.1 & \\
\enddata
\tablenotetext{\alpha}{One milli modulation amplitude (mma) equals 0.1\% change in intensity.}
\tablenotetext{\beta}{The 43.2\,min measurement is a harmonic of the orbital period.} 
\end{deluxetable}
\clearpage
\subsection{Pulsation Period}
We find that all the light curves show a period ($P$) in the narrow range of 1166--1290\,s with a significantly high
amplitude. They also show the first harmonic ($P$/2) of this dominant period $P$, except for the short run
on the 22nd of January 2006. We can even see the second harmonic ($P$/3) of this period $P$
in our data on the 20th, 21st, and 23rd of January 2006. We will ignore these harmonics
for now as they merely reflect the nonlinear pulse shape of the fundamental mode $P$, and
they are not linearly independent modes.
Each light curve
of \cvza\ therefore shows evidence of a single independent period $P$ besides the orbital period.
The value of this dominant independent period $P$ changes from 1166\,s to 1290\,s,
as shown in Table\,\ref{per-cvza}. The
suggestive lack of stability in the observed period $P$ rules out the possibility that it represents
the spin period of the white dwarf.
The spin period of a white dwarf in a CV has been shown to be extremely stable, and we expect the rate of
change of period with time to be $\dot{P}\leq10^{-14}$ \citep{Woodet05,Mauche06}.

        We now consider the possibility that the dominant period $P$ in each of the
light curves of \cvza\ represents a pulsation mode excited in the primary white dwarf. We find that the
observed range of periods 1166--1290\,s are consistent with nonradial g-mode pulsations in white dwarf stars.
Furthermore we observe high amplitudes in the long period
(600--1200\,s) \zzc\ stars \citep{Clemens93,Kanaanet02,Mukadamet06}, which is also
consistent with the observed amplitudes in \cvza.
The pulse shapes of the long period \zzc\ stars
are mostly non-sinusoidal and therefore result in harmonics in the DFTs, which is 
similar to the behavior shown by \cvza. We also see evidence of large amplitude modulation in the
pulsation spectra of long period \zzc\ stars even on timescales as short as
a few days \citep[e.g.][]{Kleinmanet98}; we explain the phenomenon of amplitude modulation below.

The eigenfrequencies of a pulsating white dwarf, representative of
its fundamental parameters and stellar structure, change only on evolutionary
timescales $\dot{\nu}\sim10^{-19}$\,Hz/s \citep[e.g.][]{Kepleret05}.
We can expect that accreting pulsating white dwarfs evolve faster than their isolated counterparts, but
we still expect changes in the eigenfrequencies to occur on extremely slow
timescales $\dot{\nu}\sim10^{-12}$\,Hz/s \citep[see][]{Townsleyet04}.
Variable
white dwarfs exhibit only a handful of all possible eigenfrequencies at any given time.
If new eigenmodes get excited in the star and
previously excited eigenmodes are entirely suppressed, then the observed pulsation spectrum
will show different frequencies.
This does not imply that the
eigenfrequencies
themselves are changing, but that the amplitude of excitation can vary on short timescales from zero to a finite value.
This amplitude modulation explains why we observe a different dominant period $P$ in each of the light curves of \cvza.
The behavior of the optical variations in \cvza\ is completely consistent with nonradial white dwarf pulsations.

\section{New accreting pulsating white dwarf \cvzb}
\begin{figure}[!ht]
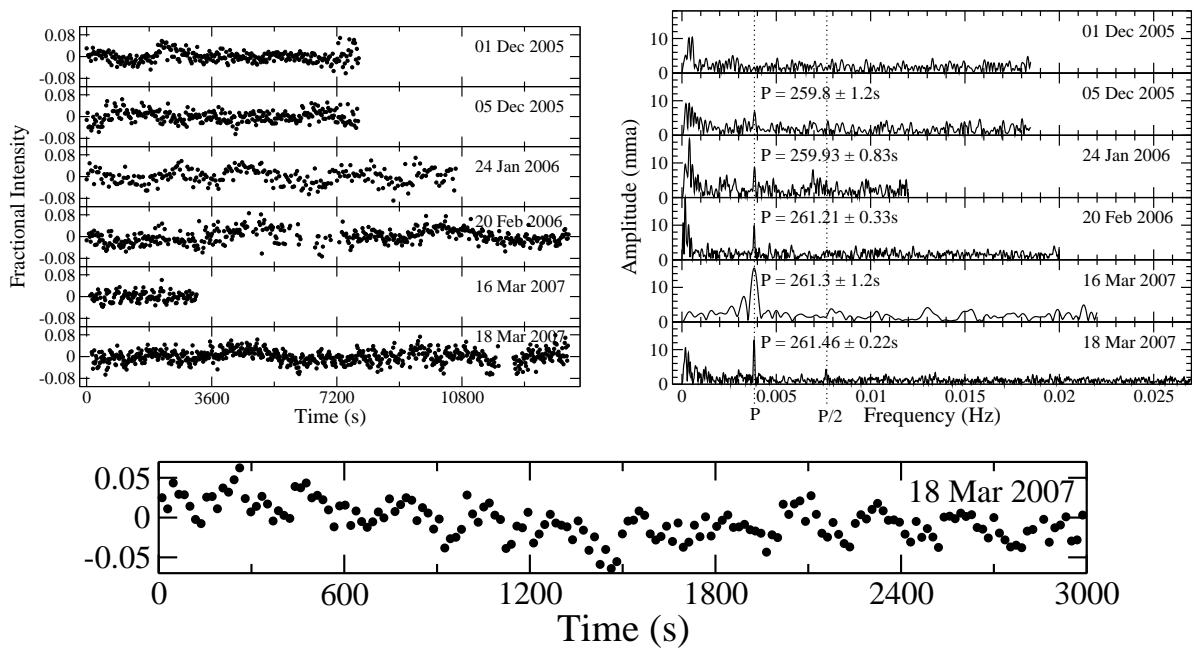

\centering
\includegraphics[width=3in,clip=true]{f3a}\hspace{0.2in}\includegraphics[width=3in,clip=true]{f3b}

\vspace{0.15in}
\includegraphics[height=1in,clip=true]{f3c}
\caption{\label{lcft-cvzb} Light curves and DFTs of \cvzb\ data acquired using the 3.5\,m telescope at APO and
the 2.2\,m telescope at CAO.
We indicate the location of the period $\sim$260\,s and its first harmonic (P/2)
against the DFTs computed for all the runs. Pulsations are clearly evident in the light curve from 18th of March 2007
shown separately.}
\end{figure}

We estimate an effective temperature of 13000\,K
for \cvzb\ ($g$\,=\,18.2), uncertain by a few 1000\,K (see Table\,\ref{t-fitparameter}).
We show the optical light curves and corresponding DFTs of \cvzb\ acquired using the
3.5\,m telescope at APO and the 2.2\,m telescope at CAO in Figure\,\ref{lcft-cvzb}.
We indicate the periods and amplitudes
of the optical variations observed on individual nights in Table\,\ref{per-cvzb}.

\clearpage                                                                                                    
\clearpage                                                                                                             
\subsection{Orbital Period}
\citet{Szkodyet05} determine the orbital period of \cvzb\ to be 84\,$\pm\,8$\,min from radial velocity measurements.
The light curve variations longer than 30\,min are all suggestive of being either direct
measurements or harmonics of this period.
Most of these measurements are consistent with each other, considering that the uncertainties
are under-estimated.
We compute a weighted average for the orbital period to be 90.9\,$\pm\,1.8$\,min,
where the weights are inversely proportional to the 1\,$\sigma$ uncertainties.
We boost the error bar to 7.3\,min to make the average value consistent with all the different measurements.
This step is necessary to derive a realistic estimate of the uncertainty in measuring the orbital period; the
observing runs are too short to determine the orbital period reliably and hence the least squares uncertainties are
severely under-estimated.
Our determination of the orbital period then becomes 91\,$\pm\,7$\,min, which is about seven minutes
longer than the determination of \citet{Szkodyet05}, but consistent
within the uncertainties.
           
\clearpage                                                                                                  
\begin{deluxetable}{lllllll}
\rotate
\tabletypesize{\footnotesize}
\tablecolumns{7}
\tablewidth{0pc}
\tablecaption{\label{per-cvzb} We list all the periodicities observed in the light curves of \cvzb.
We indicate the short period close to 260\,s in boldface,
that which we deduce to be the nonradial g-mode exhibited by the primary white dwarf.} \tablehead{
\colhead{\hspace{-0.1in}Observation} & \multicolumn{3}{c}{Orbital Period (min); Amplitude (mma)} & \multicolumn{3}{c}{Pulsation Related Period (s); Amplitude (mma)} \\
\colhead{\hspace{-0.1in}Date (UTC)}& \colhead{\hspace{-0.1in}Fundamental\,P} & \colhead{\hspace{-0.1in}First Harmonic\,P/2} & \colhead{\hspace{-0.1in}Second Harmonic\,P/3} & \colhead{\hspace{-0.1in}Fundamental\,P} & \colhead{\hspace{-0.1in}First Harmonic\,P/2} & \colhead{\hspace{-0.1in}Linear Combination$??$}}
\startdata
\hspace{-0.1in}01\,Dec\,2005 &\hspace{-0.1in} & \hspace{-0.1in}45.1\,$\pm$\,1.6; 9.1\,$\pm$\,1.6 & \hspace{-0.1in}31.1\,$\pm$\,0.8; 8.6\,$\pm$\,1.5 &\hspace{-0.1in} &\hspace{-0.1in} &\hspace{-0.1in}\\
\hspace{-0.1in}05\,Dec\,2005 & \hspace{-0.1in}&\hspace{-0.1in} &\hspace{-0.1in} &\hspace{-0.1in}{\bf 259.8\,$\pm$\,1.2; 6.8\,$\pm$\,1.7} & \hspace{-0.1in} &\hspace{-0.1in}\\
\hspace{-0.1in}24\,Jan\,2006 & \hspace{-0.1in} &\hspace{-0.1in}41.8\,$\pm$\,0.7; 17.0\,$\pm$\,2.0 &\hspace{-0.1in} &\hspace{-0.1in}{\bf 259.93\,$\pm$\,0.83; 8.6\,$\pm$\,2.0 } &\hspace{-0.1in} & \hspace{-0.1in}144.09\,$\pm$\,0.27; 8.0\,$\pm$\,2.0\tablenotemark{\gamma}\\
\hspace{-0.1in}20\,Feb\,2006 & \hspace{-0.1in}94.5\,$\pm$\,1.4; 18.6\,$\pm$\,1.2 & \hspace{-0.1in}47.0\,$\pm$\,0.5; 12.1\,$\pm$\,1.2 &\hspace{-0.1in} & \hspace{-0.1in}{\bf 261.21\,$\pm$\,0.33; 9.6\,$\pm$\,1.2}& \hspace{-0.1in} &\hspace{-0.1in}\\
\hspace{-0.1in}16\,Mar\,2007 & \hspace{-0.1in}& \hspace{-0.1in}&\hspace{-0.1in} & \hspace{-0.1in}{\bf 261.3\,$\pm$\,1.2; 15.8\,$\pm$\,1.6}& \hspace{-0.1in} &\hspace{-0.1in}\\
\hspace{-0.1in}18\,Mar\,2007 & \hspace{-0.1in}88.2\,$\pm$\,1.6; 11.4\,$\pm$\,1.0 & \hspace{-0.1in}48.8\,$\pm$\,0.6; 9.7\,$\pm$\,1.0 &\hspace{-0.1in} & \hspace{-0.1in}{\bf 261.46\,$\pm$\,0.22; 12.7\,$\pm$\,1.0}& \hspace{-0.1in}130.93\,$\pm$\,0.17; 4.2\,$\pm$\,1.0 &\hspace{-0.1in}\\
\enddata
\tablenotetext{\gamma}{This period is only a 2.5\,$\sigma$ detection.}
\end{deluxetable}

\clearpage

\subsection{Pulsation Period}
All, but one, of the light curves indicate a short period $P$ near 260\,s,
as shown in Table\,\ref{per-cvzb}.
Measurements from the 5th of December 2005 and the 24th of January 2006 indicate a value
near 259.9\,s, while the other measurements indicate a distinct value near 261.3s.
This suggests that the 
260\,s period is a doublet; sampling different
phases of the beat cycle where one mode may dominate over the other can explain the slightly
different period measurements.
In the short period \zzc\ stars (200--300\,s) close to the blue edge of the
non-interacting \zzc\ instability strip,
we often observe multiplet structure in the excited modes due to
rotation or magnetic field. These stars typically show lower amplitudes in the range
of 1--20\,mma \citep{Clemens93,Kanaanet02,Mukadamet06}, which is
consistent with the amplitudes observed for the 260\,s period.

Our present constraint on the splitting of the 260\,s mode
is about 20.6\,$\mu$Hz. If the splitting were caused by rotation, this would imply a spin
period of $\sim$1.8\,days, which is too slow for an accreting
white dwarf. It is therefore much more likely
that the splitting is caused by a magnetic field $\sim10^5$\,G \citep[see][]{Joneset89}.
We can justify that the
mode was not observed on the 1st of December 2005 due to beating of the closely spaced frequencies.
The amplitude modulation from
6.8\,mma\footnote{One milli modulation amplitude (mma) equals 0.1\% change in intensity.} on the
5th of December 2005 to 15.8\,mma on the 16th of March 2007 lends support to this idea.
Although we find no evidence to contradict our idea that the 260\,s period is a nonradial g-mode,
we are unable to eliminate the remote possibility that it could be the spin period of the white dwarf.

We expect to measure the spin period of a rapidly rotating white dwarf in a CV whenever there are
features on the surface of the white dwarf, such as a hot spot. Such a hot spot 
typically arises in accreting white dwarfs with a strong magnetic field, where the accretion
flow close to the white dwarf becomes field-aligned. The accretion stream then funnels to one
or both magnetic poles.
The stream undergoes a shock front and radiates X-rays
which heat the white dwarf surface near the pole(s). There are two kinds of CVs that show
hot spots: Intermediate Polars with magnetic fields of 1--10\,MG,
usually with some outer accretion disk, and Polars with fields $>$\,15 MG devoid of a disk.
Since the optical spectrum of \cvzb\
shows double-peaked lines, typical of an accretion disk, it is not likely to be a Polar CV.
The spectrum does not show any signs of Zeeman splitting, suggesting that the field is smaller
than 1\,MG \citep[see][]{WickramasingheaFerrario00}. Also, there is no X-ray source associated
with \cvzb\ \citep[see][and references therein]{Szkodyet05}. These constraints reduce the possibility
that \cvzb\ is an Intermediate Polar, and the 260\,s period represents the spin period of the white dwarf
caused by a rotating hot spot. However we are not completely certain because \citet{Araujo-Betancoret05}
claim to observe the spin period of 1.12\,min in the recently discovered accreting
pulsating white dwarf HS\,2331+3905; its optical spectrum does not show any Zeeman splitting
and it is not associated with any X-ray source either.

We had initially dismissed the 144\,s period observed on the 24th of January 2006 as noise due to flickering.
But we find that this period is perhaps revealing the interaction of the orbital period with the 260\,s period.
Let $\Omega$ be the frequency associated with the harmonic of the orbital period 41.8\,min 
observed on the 24th of January 2006. Let
$f_1$ and $f_2$ be the frequencies associated with the 259.93\,s and 144.09\,s periods respectively. We find
that within uncertainties $f_2\,=\,2\,(f_1-\Omega)$. This implies that $f_2$ could be a linear combination frequency,
however its amplitude is only 2.5 times the average noise amplitude, making it less reliable than a 3\,$\sigma$ detection.
We cannot explain why we do not observe $(f_1-\Omega)$, and why the 144s period is only visible
during one set of observations, if it is indeed real.
This does not help us resolve the dilemma of whether the 260\,s period represents the spin period or
a pulsation mode, as both these phenomena could potentially interact with the orbital period through tides.

Our observations on the 1st of December 2005 do not show the 260\,s period inspite of a
suitably low noise level (see Figure\,\ref{lcft-cvzb}). It is possible that the accretion is clumpy at these low
rates of mass transfer. If the accretion stopped for some time, the hot spot could cool off, explaining
the absence of the 260\,s period during our first run under this scenario. We can explain the
observed 260\,s period with two different models: nonradial pulsation and rotation. We adopt the model
of nonradial pulsation here because it is more likely.

\section{Non-variable from our search \nova}
\begin{figure}[!ht]
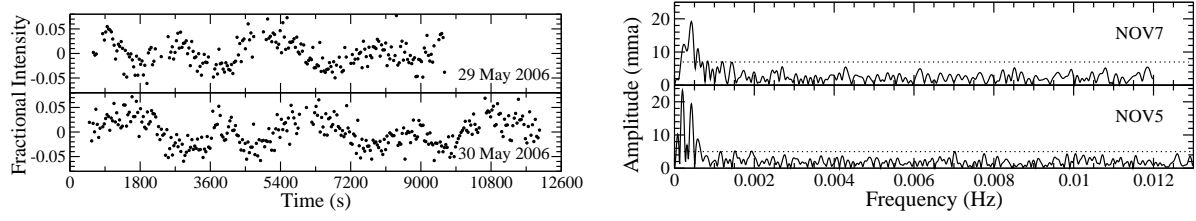

\centering
\includegraphics[width=3.0in,clip=true]{f4a}\hspace{0.2in}\includegraphics[width=3.0in,clip=true]{f4b}
\caption{\label{lcft-nova} We show the light curves and corresponding DFTs of \nova\ data obtained using the 3.5\,m WIYN telescope.
We also indicate the pulsation related non-variability limit in mma, subsequent to the NOV designation.}
\end{figure}

We report a non-variable \nova\ ($g$\,=\,20.25),
which we observed twice during our search, obtaining useful non-variability limits related
to the absence of pulsations in the primary white dwarf. Note that even CVs that show variability
from flickering, spin or orbital period modulations, would still count as non-variables in this context.
We show the light curves and DFTs of \nova\ acquired using the WIYN
3.5\,m telescope in Figure\,\ref{lcft-nova}.
We indicate the periods and amplitudes
of the optical variations related to the orbital period in Table\,\ref{per-nova}.

\clearpage

\clearpage

\begin{deluxetable}{llll}
\tabletypesize{\footnotesize}
\tablecolumns{4}
\tablewidth{0pc}
\tablecaption{\label{per-nova} We list the periods observed in the light curves of \nova, representative
of modulations related to the orbital period. We also list the non-variability limit related to pulsation.}\tablehead{
\colhead{Observation} & \multicolumn{2}{c}{Orbital Period (min); Amplitude (mma)} & \colhead{Non-variability Limit for Nonradial Pulsations} \\
\colhead{Date (UTC)}& \colhead{Fundamental\,P} & \colhead{First Harmonic\,P/2} & \colhead{}}
\startdata
29 May 2006 & & 39.4\,$\pm$\,0.7; 18.1\,$\pm$\,2.1 & $\leq$7\,mma for periods within 83--1430\,s (NOV7) \\
30 May 2006 & 83.3\,$\pm$\,1.7; 22.0\,$\pm$\,1.6 & 38.8\,$\pm$\,0.4; 18.5\,$\pm$\,1.6 & $\leq$5\,mma for periods within 77--1470\,s (NOV5) \\
\enddata
\end{deluxetable}
\clearpage

\subsection{Orbital Period}
The DFT of \nova\ acquired on the 29th of May 2006 exhibits substantial power at low frequencies, and
we fit the highest unresolved peak to determine the harmonic of the orbital period $\approx$39.4\,min. The light curve
from the next night distinctly shows both the orbital period and its harmonic at 83.3\,min \& 38.8\,min respectively.
Our weighted average of these three measurements is 78.8\,$\pm$\,1.4\,min, using weights that are inversely proportional to the 1\,$\sigma$
uncertainties. We artificially boost the error bar to 4.5\,min for consistency with all the measurements. Hence our
determination of the orbital period becomes 79\,$\pm$\,5\,min. Note that both light curves from Figure\,\ref{lcft-nova} show
a distinctive feature: the orbital modulation consists of a $\sim$30--35\,min lower amplitude feature followed by a $\sim$50\,min
relatively higher amplitude feature. 

\subsection{Pulsation related Non-variability Limit}
\nova\ was
Not Observed to Vary (NOV), and we designate it as NOV$x$, where $x$ represents
the non-variability limit determined from the DFT, similar to the scheme for non-variables within and around
the \zzc\ instability strip \citep[see][]{Mukadamet04a}.
The highest peak in the DFT essentially
defines the detection threshold or the non-variability limit, provided it is not
related to the orbital period or also present
in the DFTs of the reference stars.
In that case, we apply the same test to the second-highest peak, and so on, until we can
determine the highest peak, truly representative of the non-variability limit.

The data acquired using the WIYN 3.5\,m telescope on the 29th of May 2006 (top right panel, Figure\,\ref{lcft-nova})
allow us to constrain the pulsation amplitude below 7\,mma for
periods in the range of 83--1430\,s. The data from the 30th of May 2006 (bottom right panel, Figure\,\ref{lcft-nova})
restrict the pulsation amplitude to below 5\,mma for periods in the range of 77--1470\,s.
We conclude that there are
no pulsation related periodicities in the range of about 85--1400\,s that are higher than the limiting amplitude
of 0.5\%; we designate this non-variable as NOV5.

We estimate an effective temperature of 10500\,K
for the accreting primary white dwarf in \nova\ using its optical
SDSS spectrum (see Table\,\ref{t-fitparameter}).
Since most of the known accreting white dwarf pulsators cluster in the temperature range
of 14500--15000\,K \citep{Szkodyet07}, we wonder if the non-variable \nova\ is too cool to lie
within the instability strip(s) for variable accretors. \citet{Araujo-Betancoret05} determine an effective temperature
of 10500\,K for the accreting white dwarf pulsator HS\,2331+3905. However given the uncertainty
of $\sim$2000\,K in our temperature estimate, \nova\ could easily be substantially
cooler than HS\,2331+3905.

\section{Summary}
The light curves of \cvza, \cvzb, and \nova\ show optical variations indicative of the orbital
period of these binary systems. We determine an orbital period of
$86.3\,\pm\,2.0$\,min for \cvza,
$91\,\pm\,7$\,min for \cvzb, and $79\,\pm\,5$\,min for \nova\ using our data.

We also observe a single independent period in the 
light curves of \cvza\ in the range of 1166--1290\,s, which varies from run to run.
The spin of the white dwarf is expected to be highly stable and cannot explain the observed variations.
We deduce that the variations in the light curves of \cvza\ with long periods, high amplitudes of 4.5--7\%, non-sinusoidal
pulse shapes,
and substantial amplitude modulation are completely consistent 
with nonradial g-modes, similar to those observed for the non-interacting long period cool \zzc\ stars. 

	We observe a persistent period close to 260\,s in five out of six of the light curves of \cvzb.
We adopt the explanation that this period is a nonradial pulsation mode with doublet structure
and consequent beating, similar to the modes excited in the short period \zzc\ stars. The observed
amplitudes are also consistent with the small amplitudes observed in the short period \zzc\ stars, found close
to the blue edge of the \zzc\ instability strip.
But we cannot rule out the remote possibility
that the 260\,s period could be the spin period of the white dwarf without additional data that span
a long timebase.

Even after observing \nova\ twice, we did not find 
any pulsation related periodicities in the range of 85--1400\,s that are higher than the limiting amplitude
of 0.5\% (NOV5). We therefore conclude that \nova\ does not exhibit nonradial pulsations.
Our temperature estimate of 10500\,K for this accreting white dwarf suggests that it
may perhaps be too cool to pulsate. This discovery paper brings the 
total number of known accreting white dwarf pulsators to eleven.

\acknowledgments
Support for this work was provided by NASA through the Hubble Fellowship grant
HST-HF-01175.01-A awarded by the Space Telescope Science Institute, which is operated
by the Association of Universities for Research in Astronomy, Inc., for NASA, under contract NAS 5-26555.
We thank NSF for the grant AST-0607840.
BTG was supported by a PPARC Advanced
Fellowship. AA thanks the Royal Thai Government for a studentship.
Based on observations obtained with the Apache Point Observatory 3.5-meter telescope, which is owned and operated by the Astrophysical Research Consortium and
on observations collected at the Centro Astron\'{o}mico Hispano Alem\'{a}n (CAHA) at Calar Alto,
operated jointly by the Max-Planck Institut f\"{u}r Astronomie and the Instituto de Astrof\'{i}sica de
Andaluc\'{i}a (CSIC). Also based on observations obtained at the
WIYN Observatory, which is a joint facility of the University of Wisconsin-Madison, Indiana University, Yale University, and the National Optical Astronomy Observatories.
We thank the referee for helping us make this manuscript a better paper.

\end{document}